\documentclass[cits]{PoS}
\usepackage[utf8]{inputenc}
\usepackage{verbatim}

\title{The Aoki phase revisited}

\ShortTitle{The Aoki phase revisited}

\author{\speaker{Alejandro Vaquero}\thanks{The speaker wants to thank M. Lüscher for his help in implementing the SAP preconditioner used in this work.}\\
	Departamento de Fisica Teorica, Universidad de Zaragoza\\
        Cl Pedro Cerbuna 12, E-50009 Zaragoza (Spain)\\
	E-mail: \email{alexv@unizar.es}}

\author{Vicente Azcoiti\\
	Departamento de Fisica Teorica, Universidad de Zaragoza\\
        Cl Pedro Cerbuna 12, E-50009 Zaragoza (Spain)\\
	E-mail: \email{azcoiti@azcoiti.unizar.es}}

\author{Giuseppe Di Carlo\\
	INFN, Laboratori Nazionali del Gran Sasso,\\
	I-67100 Assergi-L'Aquila (Italy)\\
	E-mail: \email{gdicarlo@lngs.infn.it}}

\author{Eduardo Follana\\
	Departamento de Fisica Teorica, Universidad de Zaragoza\\
        Cl Pedro Cerbuna 12, E-50009 Zaragoza (Spain)\\
	E-mail: \email{efollana@unizar.es}}

\abstract{In order to elucidate the vacuum structure of the Aoki phase, we carried out a numerical investigation of QCD with two flavours of Wilson fermions, within the p.d.f. framework and \emph{in the absence of external sources}. The simulations performed at $V=4^4$ suggest a rich vacuum structure, where the observable $i\bar\psi\gamma_5\psi$ is allowed to take non-zero values of the same order of magnitude than the order parameter of the Aoki phase $i\bar\psi\gamma_5\tau_3\psi$. However, the simulation at higher volumes $V=6^4$ suffers from large statistical errors.}

\FullConference{The XXVIII International Symposium on Lattice Field Theory\\
		June 14-19, 2010\\
		Villasimius, Italy}

\begin{document}

\section{Introduction}

A year ago, we applied the p.d.f. formalism \cite{pdf} to QCD with 2 flavours of
Wilson fermions, and in particular to the Aoki phase \cite{Aoki}, to analyze its 
spontaneous flavour and Parity breaking pattern \cite{yo2}. The investigation led 
us to several interesting and undocumented results, which depended on the properties 
of the spectral density $\rho_U\left(\mu\right)$ of the Hermitian Dirac Wilson operator 
$H = \gamma_5\Delta$. The spectral density should be an even function of the eigenvalues 
in the continuum limit $a\rightarrow 0$, but it has been also suggested that 
it may become symmetric already in the thermodynamic limit $V\rightarrow\infty$. 
These two different behaviours come out into different possibilities:

\begin{enumerate}
\item If the spectral density becomes an even function of the eigenvalues \emph{only} in the 
continuum limit, then violations of hermiticity are expected to
induce large artifacts in the measurement 
of the $\eta-$meson mass at finite lattice spacing, even if $V\rightarrow\infty$. 
\item If, on the contrary, it is enough to reach the thermodynamic limit to find a symmetric 
spectral density, that is the most plausible case \cite{yo2,Sharpe}, then two new scenarios 
appear as $V\rightarrow\infty$, where the Aoki phase 
displays a rich, unexpected behaviour. In the first scenario, the conditions
\begin{equation}
\left\langle\left(i\bar\psi\gamma_5\psi\right)^{2n}\right\rangle \neq 0 \qquad n\in\mathbb{N}
\label{Aoki-PoS-I}
\end{equation}
are verified as a consequence of the spectral symmetry \cite{yo2}. The standard 
picture of the 
Aoki phase is incomplete, and new phases, unrelated to the original Aoki phase, appear. 
The second possibility, predicted by the chiral lagrangians \cite{Sharpe2,Sharpe}, complies 
with the standard picture of the Aoki phase. The following equations
\begin{equation}
\left\langle\left(i\bar\psi\gamma_5\psi\right)^{2n}\right\rangle = 0 \qquad n\in\mathbb{N}
\label{Aoki-PoS-II}
\end{equation}
are satisfied, and this give rise to an infinite set of independent sum rules for the 
eigenvalues $\mu$, one for each $n$. These scenarios are mutually exclusive. 
\end{enumerate}
Up to now, there is no theoretical proof selecting one of these possibilities. 
In \cite{yo2}, an antisymmetric spectral density in the thermodynamic limit was excluded,
 based on quenched simulations. 
Therefore we decided to perform the dynamical fermions simulations required to distinguish between 
(\ref{Aoki-PoS-I}) and (\ref{Aoki-PoS-II}).

The current paper is organized as follows: In the next section, the details and technical 
difficulties of the simulation of dynamical fermions inside the Aoki phase are reviewed. 
Section 3 shows and analyses our numerical results, and the last section is devoted to our 
final conclusions.

\section{Simulating the Aoki phase without external sources}

Although the simulations of the Aoki phase with dynamical fermions are nothing new in the 
lattice QCD panorama, these have always been performed under very special conditions: 
an external source is added to the action with a twisted mass term

\begin{equation}
ih\bar\psi\gamma_5\tau_3\psi
\label{Aoki-PoS-III}
\end{equation}
in order to (i.) regularise the small eigenvalues of the Dirac Wilson operator, which 
appear only in the Aoki phase and usually spoil any attempt of simulation without the 
external source, and (ii.) to analyse the pattern of spontaneous flavour and Parity 
breaking. As proved in \cite{yo2}, the use of an external source like (\ref{Aoki-PoS-III}) 
selects a standard Aoki vacuum, complying with (\ref{Aoki-PoS-II}). The only way to 
investigate the existence of the new phases characterized by (\ref{Aoki-PoS-I}) is to 
remove the external source, and extract results from direct measurements of the Gibbs 
state ($\epsilon$-regime). The latter point is solved by the p.d.f. formalism, but the former 
--the removal 
of the external source-- has been an unexplored option in the Aoki phase dynamical simulations 
for a number of years. The reason is the appearance of small eigenvalues of the Hermitian Dirac 
Wilson operator, of order $O\left(\frac{1}{V}\right)$. As far as we know, the technical 
problems inside the Aoki phase are similar to those faced when trying to reach the physical 
point: The critical slowing down spoils the efficiency of simulations, the Dirac Wilson 
operator becomes increasingly harder to invert, and the performance decreases dramatically. 
In fact, for some values of the parameters $\left(\beta,\kappa\right)$ in the coupling-mass 
phase diagram, the standard algorithms to invert the Dirac operator just will not work. 
This fact called for a research on competent algorithms to simulate inside the Aoki phase.

Inspiration came in the recent results for new algorithms which reduce the critical slowing 
down for small masses in several orders of magnitude. In this work, we successfully 
implemented a SAP preconditioner in a GCR inverter, as explained in \cite{Luescher}. 
The new inverter allow us to perform simulations inside the Aoki phase without external 
sources at a reasonable speed. Unfortunately this is not the whole history.

The simulations done outside the Aoki phase feature a spectral gap around the origin, 
with a symmetric spectrum\footnote{By symmetric {\bf \emph{here}} we mean that the 
number of positive and negative eigenvalues are the same.}. This gap is required to 
preserve Parity and flavour symmetry \cite{yo2}. However, the Aoki phase breaks both 
of them, so this gap was not present in our simulations, and as explained above, the 
smallest eigenvalues of each configuration took values of order $O\left(\frac{1}{V}\right)$. 
As the eigenvalues approach the origin, it may happen that they try to change sign, 
rendering the spectrum asymmetric, but this movement is forbidden by the 
Hybrid Montecarlo dynamics, i.e., \emph{the HMC algorithm is not ergodic for 
Wilson fermions inside the Aoki phase}. Therefore we are introducing artificial 
constraints in the dynamics, and the final results are bound to be modified. 
That is why we considered another dynamical fermion simulation algorithm, the 
Microcanonical Fermionic Average (MFA) algorithm \cite{MFA}, which solves the 
problem of the eigenvalue crossing, but converges poorly as the volume increases.

The problem was overcome by using an argument developed in \cite{QCharge} and 
applied to the current case in \cite{Sharpe}, where the asymmetry of the 
spectrum $n_{Asym}$ is related to the topological charge $Q$ as

\begin{equation}
n_{Asym}\propto Q.
\label{Aoki-PoS-IV}
\end{equation}
Then the constraints imposed by the hybrid Montecarlo are equivalent to leaving 
the topological charge fixed. Since the measurement of the observables should 
not depend on the value of the topological charge in the thermodynamic limit, 
we can select the symmetric sector $Q=0$ and measure our observables there, 
where one expects to have smaller finite volume effects.

Another interesting possibility is (i.) to measure the weight of the different $n_{Asym}$ 
sectors in the partition function via the MFA algorithm, (ii.) then perform Hybrid 
Montecarlo simulations within the relevant sectors, and (iii.) do a weighted average 
of the observables obtained in the HMC using the MFA weights. The results obtained 
with this method were contrasted to those coming from the direct MFA simulations, 
and with the HMC simulations at fixed $n_{Asym}$.

A third proposal for simulations was announced during the talk: The addition of 
a small (\ref{Aoki-PoS-III}) external source, large enough to regularise the small 
eigenvalues, but small enough to avoid vacuum selection, would allow us to include 
the eigenvalue crossing phenomenon in our HMC algoritm. Unfortunately, there exists 
no value of the external field $h$ capable of this two deeds at the same time. If the 
field is very small, no vacuum is selected, but the eigenvalues do not cross the 
origin during the simulations; on the contrary, for larger values of the field $h$, 
the standard Aoki vacuum is selected. This third possibility was, thus, forsaken.

\section{Numerical results}

We performed measurements of three observables of interest
\begin{table}[hc]
\caption{Expected behaviour of the analysed observables in the different 
scenarios as $V\rightarrow\infty$.}
\begin{center}
\begin{tabular}{|c|c|c|c|}
\hline
 & \begin{tabular}{c} Outside\\Aoki\end{tabular} & \begin{tabular}{c} 
Aoki\\Standard Wisdom\end{tabular} & \begin{tabular}{c} Aoki\\Our Proposal
\end{tabular} \\
\hline
$\left\langle\left(i\bar\psi_u\gamma_5\psi_u\right)^{2}\right\rangle$ & $\sim 0$ 
& $\neq 0$ & $\neq 0$ \\
$\left\langle\left(i\bar\psi\gamma_5\psi\right)^{2}\right\rangle$ & $\sim 0$ 
& $\sim 0$ & $\neq 0$ \\
$\left\langle\left(i\bar\psi\gamma_5\tau_3\psi\right)^{2}\right\rangle$ 
& $\sim 0$ & $\neq 0$ & $\neq 0$ \\
\hline
\end{tabular}
\end{center}
\end{table}

Our measurements refer always to the second moment of the p.d.f., which 
should be zero in case of symmetry conservation, and non-zero if the symmetry 
is spontaneously broken \cite{pdf}. The first observable signals Parity
breaking, and should be non-zero inside the Aoki phase. The second one allows us 
to distinguish between our proposal --there is an additional \emph{Aoki-like phase} 
verifying $\left\langle\left(i\bar\psi\gamma_5\psi\right)^{2}\right\rangle\neq 0$-- 
and the one involving an infinite number of sum rules. Finally, the third observable 
is the landmark of the Aoki phase, marking spontaneous flavour and Parity breaking.

The first set of simulations were performed using the HMC algorithm, improved with a 
SAP-preconditioned solver. The symmetric runs $n_{Asym} = 0$ were performed starting 
from a cold (ordered, links close to the identity) and from a hot (disordered, close 
to strong coupling) configuration, obtaining identical results within errors. We could 
not find an asymmetric state $n_{Asym} = 1$ in a cold configuration at the values 
of $\kappa$ explored, the asymmetric run was started only from a hot configuration.

\begin{table}[hc]
\caption{Results of the Hybrid Montecarlo measurements.}
\begin{center}
\begin{tabular}{|c|c|c|c|c|}
\hline
 & \begin{tabular}{c}HMC $V=4^4$\\Outside Aoki$^\star$\end{tabular} & 
\begin{tabular}{c} HMC$^{\star\star}$ $V=4^4$\\$n_{Asym}=0$\end{tabular} & 
\begin{tabular}{c} HMC $V=6^4$\\$n_{Asym}=0$\end{tabular} & \begin{tabular}{c} HMC $V=4^4$\\
$n_{Asym}=1$\end{tabular} \\
\hline
NConf & 20002 & 19673 & 664 & 10002 \\
\hline
$\left\langle\left(i\bar\psi_u\gamma_5\psi_u\right)^2\right\rangle$ & $2.098(3)
\times 10^{-3}$ & $1.90(3)\times 10^{-2}$ & $9.2(30)\times 10^{-3}$ & $6.52(72)
\times 10^{-3}$ \\
$\left\langle\left(i\bar\psi\gamma_5\psi\right)^2\right\rangle$ & $4.149(4)
\times 10^{-3}$ & $2.69(12)\times 10^{-2}$ & $-4.3(30)\times 10^{-1}$ & $-4.49(50)
\times 10^{-2}$ \\
$\left\langle\left(i\bar\psi\gamma_5\tau_3\psi\right)^2\right\rangle$ & $4.244(4)
\times 10^{-3}$ & $4.90(9)\times 10^{-2}$ & $4.7(30)\times 10^{-1}$ & $7.10(30)
\times 10^{-2}$ \\
\hline
\multicolumn{5}{l}{\footnotesize $^\star$Point outside the Aoki phase $\beta=3.0$, 
$\kappa=0.22$.} \\
\multicolumn{5}{l}{\footnotesize $^{\star\star}$Point inside the Aoki phase $\beta=2.0$, 
$\kappa=0.25$.}
\end{tabular}
\end{center}
\end{table}

We expect all the observables to have non-zero expectation values, even outside the 
Aoki phase, due to finite volume effects. However, the values inside the Aoki phase 
are an order of magnitude larger than those measured outside the Aoki phase. Outside 
the Aoki phase, the following approximate rule holds:

$$2\left\langle\left(i\bar\psi_u\gamma_5\psi_u\right)^2\right\rangle\approx
\left\langle\left(i\bar\psi\gamma_5\psi\right)^2\right\rangle\approx\left\langle
\left(i\bar\psi\gamma_5\tau_3\psi\right)^2\right\rangle,$$
which is a manifestation of (i.) the presence of an spectral gap and (ii.) the high 
level of symmetry of the spectral density, even at small volumes. This facts can 
be seen in the p.d.f. expressions for these observables in terms of the eigenvalues $\mu$

\begin{eqnarray}
\left\langle\left(i\bar\psi_u\gamma_5\psi_u\right)^2\right\rangle & = & 
\frac{1}{V^2}\left[\sum_{j}^N\left(\frac{1}{\mu_j^2}\right) - \left(\sum_{j}^N
\frac{1}{\mu_j}\right)^2\right],\\
\left\langle\left(i\bar\psi\gamma_5\psi\right)^2\right\rangle & = & \frac{1}{V^2}
\left[2\sum_{j}^N\left(\frac{1}{\mu_j^2}\right) - 4\left(\sum_{j}^N\frac{1}{\mu_j}
\right)^2\right],\\
\left\langle\left(i\bar\psi\gamma_5\tau_3\psi\right)^2\right\rangle & = & 
\frac{2}{V^2}\sum_{j}^N\left(\frac{1}{\mu_j^2}\right),
\end{eqnarray}
where the term $\left(\sum_{j}^N\frac{1}{\mu_j}\right)^2$ almost vanishes outside 
the Aoki phase. Nonetheless, inside the Aoki phase the gap disappears,
the asymmetry near to the origin can become relevant and this 
rule would break down.

The numerical results are reported in Table 2. Some comments are in order:
inside the Aoki phase, due to the parity and flavour breaking, we expect to 
find  $\left(i\bar\psi_u\gamma_5\psi_u\right)^2$ and  $\left(i\bar\psi\gamma_5
\tau_3\psi\right)^2$ to be large with respect to the case of standard QCD
(outside the Aoki phase), and this is exactely what we get. The crucial point
to discriminate between the two scenarios depicted above is the behaviour of
$\left(i\bar\psi\gamma_5\psi\right)^2$: if we take the results of 
$n_{Asym}=0$ in the $4^4$ lattice at face value
we are induced to conclude that the non-standard scenario for the Aoki
phase is favoured, being the $\left(i\bar\psi\gamma_5\psi\right)^2$
expectation value of the same order of magnitude of the other two observables
(and an order of magnitude larger than outside the Aoki phase). The
results of the $6^4$ lattice seem to add no useful informations due to the
large statistical errors (we hope to get better quality results in
the future).
On the other hand the result for $n_{Asym}=1$ can seem strange at a first sight
(a negative number for the expectation value of the square of an hermitian operator),
but we have to take into account that we are restricting ourselves to a single 
topological sector and what is relevant is the relative weight of the various
topological sectors to the final result (that should be positive).
The evaluation of the weights can not be performed using HMC simulations
due to ergodicity problems. 


We also observed that, in order to 
achieve in the asymmetric run the same acceptance ratios ($\sim 90\%$) than 
in the symmetric runs, we had to reduce the simulation step by a factor of 
ten, for the forces inside the HMC became much larger than expected. We took 
this as an indication that the system was trying to return to the symmetric 
state, pushing the eigenvalues through the origin, thus increasing in an 
uncontrolled manner the norm of the inverse of the Dirac Wilson operator. 
In the symmetric run, the eigenvalues certainly tried to cross the origin, 
however this did not happen continuously, but only from time to time. 
Hence, although we can not fully rely in any of these results because of the 
aforementioned problems of the HMC, we find the results of $n_{Asym}=0$ 
much more believable of those of the asymmetric state.

Then we introduced the MFA algorithm\cite{MFA}, with the hope of solving the 
ergodicity problems. As in the MFA the 
contributions of fermions is added during the measurement, the eigenvalues 
can cross the origin at will. 
So the MFA algorithm allowed us to measure the weights of the different sectors 
$n_{Asym}=0,1\ldots$, and then use these weights to correctly average the HMC data.



\begin{table}[hc]
\caption{Weights of the different sectors according to MFA algorithm.}
\begin{center}
\begin{tabular}{|c|c|c|c|c|}
\hline
Volume & $n_{Asym} = 0$  & $n_{Asym} = 1$ & $n_{Asym} = 2$ \\
\hline
$4^4$ & $85.9\pm5.8\%$ & $14.1\pm5.8\%$ & $0\%$ \\
\hline
\end{tabular}
\end{center}
\end{table}

\begin{table}[hc]
\caption{Results for weighted HMC.}
\begin{center}
\begin{tabular}{|c|c|}
\hline
 & \begin{tabular}{c} Weighted HMC\\$V=4^4$\end{tabular}\\
\hline
$\left\langle\left(i\bar\psi_u\gamma_5\psi_u\right)^2\right\rangle$ & $1.72(9)
\times 10^{-2}$ \\
$\left\langle\left(i\bar\psi\gamma_5\psi\right)^2\right\rangle$ & $1.7(6)
\times 10^{-2}$ \\
$\left\langle\left(i\bar\psi\gamma_5\tau_3\psi\right)^2\right\rangle$ & $5.2(3)
\times 10^{-2}$ \\
\hline
\end{tabular}
\end{center}
\end{table}

\section{Conclusions}

We have performed dynamical simulations of QCD with two flavours of Wilson fermions 
inside the Aoki phase, without any external source. We succeed overcoming the 
critical slowing-down of the simulations by using new algorithms developed recently, 
but our data showed that the HMC algorithm is not ergodic inside the Aoki phase 
without twisted mass term. The measurements collected reveal the presence of a 
Parity and flavour breaking phase through the non-zero expectation value of the 
operators $\left(i\bar\psi_u\gamma_5\psi_u\right)^2$ and $\left(i\bar\psi_u
\gamma_5\tau_3\psi_u\right)^2$, where the eigenvalues of the Dirac Wilson 
operator could take small values up to $O\left(\frac{1}{V}\right)$. 
Concerning the expectation value of $\left(i\bar\psi\gamma_5\psi\right)^2$, 
the results in the $4^4$ lattice point clearly in favour of a non-standard scenario 
for the Aoki phase, whereas the $6^4$ lattice results are not conclusive because of 
larger statistical errors.
The possibility of increasing the statistics and the volume of the lattices
used in the simulations would require large computer facilities, due to the 
high cost of the inversion of the Dirac Wilson operator in presence of 
small eigenvalues. Therefore we are working in theoretical arguments that 
may help to clarify the issue.

\section{Acknowledgments}
This work has been partially supported by an INFN-MICINN collaboration, MICINN
(grant FPA2009-09638), DGIID-DGA (grant2007-E24/2) and by the European Union under Grant Agreement number
PITN-GA-2009-238353 (ITN STRONGnet). E. Follana
is supported by MICINN through the Ramón y Cajal
program, and A. Vaquero is supported by MICINN through the FPU program.

\appendix

\section{Notation}

Even though the p.d.f. is quite a powerful tool, its use is not very 
widespread in quantum field theories. That is why the language of the 
p.d.f. is a bit different to that of QFT's. This fact sometimes leads 
to confusion and misunderstandings, therefore we want to devote a few 
lines to explain the notation used thorough this paper.

When we are referring to the observable $i\bar\psi\gamma_5\psi$, what we mean is
$$i\bar\psi\gamma_5\psi = \frac{1}{V}\sum_{x}^V i\bar\psi\left(x\right)
\gamma_5\psi\left(x\right).$$
Thence, the expectation value of the operator is computed as

$$\left\langle i\bar\psi\gamma_5\psi\right\rangle = \frac{1}{V}\left\langle
\sum_{x}^V i\bar\psi\left(x\right)\gamma_5\psi\left(x\right)\right\rangle,$$
and its higher powers become
$$\left\langle\left(i\bar\psi\gamma_5\psi\right)^k\right\rangle = \left
\langle\left(\frac{1}{V}\sum_{x}^Vi\bar\psi\left(x\right)\gamma_5\psi
\left(x\right)\right)^k\right\rangle.$$

\end{document}